\documentclass[aps,prl,twocolumn,groupedaddress,showpacs]{revtex4}
\usepackage{times,xspace}\usepackage{amsbsy,amssymb,amsmath,bm}
\usepackage{graphicx,color,epsfig,rotate}
\begin{document}

\title{Elastocaloric effect associated with the martensitic transition
in shape-memory alloys}

\author{Erell Bonnot}
\author{Ricardo Romero}
\altaffiliation{Permanent  adress:  IFIMAT, Universidad  del  Centro de
la  Provincia  de  Buenos Aires.   Pinto, 399. 7000 Tandil. Argentina.}
\author{Llu\'{i}s Ma\~nosa}
\author{Eduard Vives}
\author{Antoni Planes}
\affiliation{Departament d'Estructura i Constituents de
la  Mat\`eria.  Facultat  de  F\'\i sica.   Universitat de  Barcelona.
Diagonal, 647, E-08028 Barcelona, Catalonia.}

\date{\today}

\begin{abstract}

The elastocaloric  effect in the vicinity of the martensitic transition of a Cu-Zn-Al has been studied by inducing the transition by strain or
stress measurements. While transition trajectories show significant differences, the entropy change associated with the whole transformation
($\Delta S_t$) is coincident in both kind of experiments since entropy production is small compared to $\Delta S_t$. The values agree with
estimations based on the Clausius-Clapeyron equation. The possibility of using these materials for mechanical refrigeration is also discussed.

\end{abstract}


\maketitle

Caloric  effects are  expected to  occur under  the application  of an external  field   to  a  given  material.   The  elastocaloric  effect
\cite{piezocaloric} is  the mechanical analogue  of the magnetocaloric effect that  has received considerable  attention in the  recent years owing
to its potential  use for environmentally friendly refrigeration \cite{Tishin/Bruck2003/5}. The magnetocaloric effect is related to the isothermal
change  of entropy or  the adiabatic change  of temperature that takes place within a material when a magnetic field is applied or removed. This
effect originates from the coupling between the magnetic sublattice and an externally applied magnetic field and thus occurs in any magnetic
material.  A large effect is expected  in the vicinity of field-induced,  first-order  phase  transitions  where  large  entropy changes   should
occur   \cite{Gschneidner2005}.   By  analogy,   the elastocaloric effect is defined as the isothermal change of entropy or the adiabatic change of
temperature that takes place when a mechanical field (stress)  is applied  or released in  a given  material. Indeed, this effect is expected to be
a consequence of the coupling between an external applied stress and  the lattice. Continuing with the analogy, a large  elastocaloric effect is
also foreseen in  systems undergoing stress-induced, first-order phase transitions. Good candidates to show this  effect  are  shape-memory  alloys.
These  materials  undergo  a diffusionless  purely structural transition  from a  cubic to  a lower symmetry phase that can be stress induced
\cite{Planes2001}. Actually, shape-memory properties  are related to  this transition and  refer to the ability  of these systems  to remember their
original  shape after severe deformation \cite{Otsuka1998}.

In contrast  to magnetism, instead  of controlling the  applied stress (or  force)   which  is  the  variable thermodynamically equivalent to  the
magnetic field,  in mechanical experiments, the system is usually driven by controlling  the strain (generalized displacement) which is the
conjugated variable to the stress in the way that magnetization is the conjugated variable of  the magnetic field. In magnetic systems, due to the
difficulty in controlling magnetization, magnetization-driven experiments aimed at studying the magnetocaloric effect have not, to our knowledge,
been reported. Thus, comparing results from  both field/stress-driven and magnetization/strain-driven experiments is of general interest since
constraining   the (generalized) displacement prevents free  motion of the interfaces and therefore field/stress fluctuation can occur. Indeed, this
is especially relevant in systems undergoing a  phase transition leading to macroscopic instability. The  study of mechanical systems naturally
opens up the possibility of performing such kind of experiments. In a recent   work \cite{Bonnot2007}   we have shown that metastable trajectories
exhibit a strong dependence on the driving mechanism. In particular, strain-driven trajectories   are characterized  by the occurrence of re-entrant
behaviour  and lower dissipation than stress-driven trajectories.

The  present letter  is aimed  at studying  the elastocaloric effect in the  vicinity of the martensitic transition in a
Cu$_{68.13}$Zn$_{15.74}$Al$_{16.13}$  single  crystal (molar volume = 7.52 cm$^3$mol$^{-1}$) in  both the stress-driven and the strain-driven modes.
The sample was mechanically machined with cylindrical heads  and the  body has flat faces  35 mm long, 4  mm wide  and 1.4  mm thick. Its  axis is
close to  the [100] crystallographic direction  of  the cubic phase.  This  orientation provides  high  transformation strains for relatively  low
uniaxial applied tensile  loads along  this direction  thus  preventing  the occurrence of irreversible plastic  effects. The sample  was
conveniently heat treated to ensure that it was free from internal stresses  and  that  the order   state and vacancy concentration are close  to
their ground state values \cite{Bonnot2007}. In the absence of any applied stress, the crystal undergoes a martensitic transition from a cubic
L2$_1$ phase to  a multivariant 18R martensite at $T_M$ = 234 K.

For  strain-driven experiments we  used an  Instron 4302  screw driven tensile machine in which the  elongation is the control parameter. For
stress-driven experiments  we used a machine which  enables control of the  force applied  to the  sample while  elongation  was continuously
monitored \cite{Bonnot2007}.  The machine applies  a dead load  to the sample  which can  be  increased  or decreased  at  a well  controlled rate.
A  cryofurnace (with temperature stability $\pm$ 0.1 K) can be adapted to  both devices. All experiments have been performed at low rates [$\sim$
0.3 mm min$^{-1}$ (strain-driven) and $\sim$ 5 MPa min$^{-1}$ (stress-driven)].

Let  us now  consider a  thermodynamic system  described  by variables $\{{\bf  X},  {\bf  Y},  T   \}$,  where  $\bf  X$  is  a  generalized
displacement and  $\bf Y$ is the corresponding  conjugated field ($\bf X$  and  $\bf Y$  have  the  same tensorial  order),  and  $T$ is  the
temperature. A change in the  generalized displacement gives rise to a caloric  effect. If  this change  is induced  by an  isothermal change
$\Delta {\bf Y}$  of its conjugated field, the  caloric effect must be quantified by the corresponding  induced entropy change which is given by
\begin{equation}
\Delta   S   =  \int_{\Delta   {\bf   Y}}  \left(\frac{\partial   {\bf
X}}{\partial T} \right)_{{\bf Y}} d \bf Y,
\end{equation}
where the  generalized Maxwell relation  $(\partial S /  \partial {\bf Y})_T$ =  $(\partial {\bf X} /  \partial T)_{{\bf Y}}$  has been taken into
account.  When the field  is an uniaxial tensile  stress $\sigma$ for which  the corresponding strain (or relative  elongation along the direction
of  the  applied   force)  is  $\varepsilon$,  the  induced isothermal entropy change defining  the elastocaloric effect, is given by
\begin{equation}
\Delta S (0 \rightarrow \sigma) = \int_0^{\sigma} \left(\frac{\partial
\varepsilon}{\partial T} \right)_{\sigma} d \sigma. \label{eq1}
\end{equation}
This  expression is formally  analogous to  the expression  giving the field-induced entropy  change which defines  the magnetocaloric effect in a
magnetic system \cite{Tishin/Bruck2003/5}. If instead of $\sigma$ the   controlled  variable  is   $\varepsilon$,  the   entropy  change
corresponding  to an  isothermal variation  of  the strain  from 0  to $\varepsilon$ is given by
\begin{equation}
\Delta   S  (0   \rightarrow   \varepsilon)  =   -\int_0^{\varepsilon}
\left(\frac{\partial   \sigma}{\partial  T}   \right)_{\varepsilon}  d
\varepsilon,
\label{eq2}
\end{equation}
where the  Maxwell relation $(\partial S /  \partial \varepsilon)_T$ = $-(\partial \sigma / \partial  T)_{\varepsilon}$ has been used in this case.
Of course,  if $\varepsilon$ is the strain  corresponding to the stress  $\sigma$,  in  strict  equilibrium $\Delta  S  (0  \rightarrow \sigma) =
\Delta S (0 \rightarrow \varepsilon)$.

Now assume a  system subjected to an applied  uniaxial stress
$\sigma$ that undergoes a first-order structural (martensitic)
phase transition in equilibrium at a temperature  $T_t$. The
transition is in this case characterized  by  discontinuities in
variables  such  as strain  and entropy that  are
thermodynamically  conjugated  to  the  intensive variables stress
and temperature. In  the vicinity of  the transition the following
behaviour   of   the  strain   is   thus   expected,
$\varepsilon(T, \sigma) = \varepsilon_0  + \Delta \varepsilon \;
{\cal F}[(T_t(\sigma)- T)/\Delta  T]$, where ${\cal F}$  is a
shape-function and $\Delta  T$ is a measure  of the temperature
range  over which the transition spreads. In strict equilibrium,
$\Delta T \rightarrow 0$ so that  ${\cal  F}$  approaches   the
Heaviside  step  function.  Using expression (\ref{eq1})  and
assuming that $\varepsilon_0$ and $\Delta \varepsilon$ are
constant, in this equilibrium case the elastocaloric effect in the
vicinity of the transition is given by
\begin{equation}
    \Delta S(0 \rightarrow \sigma) = \left \{
    \begin{array}{cr}
        - \frac{\Delta \varepsilon}{\alpha} &  \qquad \mbox{for $T \in
        [T_t(0),  T_t(\sigma)]$} \\  0  & \qquad  \mbox{for $T  \notin
        [T_t(0), T_t(\sigma)]$}
    \end{array}
    \right. \, ,
\label{eq4}
\end{equation}
where $\alpha  \equiv dT_t/d\sigma$ is assumed to  be constant. Taking into  account  the  Clausius-Clapeyron  equation,  $\alpha  =  -\Delta
\varepsilon  /  \Delta S_t$,  where  $\Delta  S_t$  is the  transition entropy  change.  Therefore,  as  expected,  $\Delta  S(0  \rightarrow
\sigma) =  \Delta S_t$, and $\Delta  T = T_{\sigma} -  T_t(0) = \alpha \sigma$.  Indeed,  the same  result  is  obtained  in this  case  from Eq.
(\ref{eq2}).

Actually, these  transitions  are spread  over a  small range  of
$\sigma$, and, more importantly, they display hysteresis, which
reflects the existence of non-equilibrium dissipative effects. In
this case, taking into account the  Clausius inequality, $\oint
\delta q/T \leq 0$, for an isothermal  process, the entropy change
must satisfy,  $\Delta S(0 \rightarrow \sigma \; \text{or} \;
\varepsilon) = \frac{q}{T} + S_i$, where $S_i \geq 0$ is the
entropy production and is expected to depend on  the   actual
trajectory  followed by the   system  (strain-  or stress-driven
in our case). The preceding equation  indicates that an estimation
of  the entropy  change  based  on expression (\ref{eq1}) should
differ from an estimation  based  on direct  calorimetric
measurements. For the system of  interest here, as hysteresis is
quite small, estimations  of the entropy change based on
stress-driven and strain-driven curves should provide reasonably
good estimations of the elastocaloric effect.

\begin{figure}[t]
    \centering \includegraphics[width=0.75\linewidth,clip]{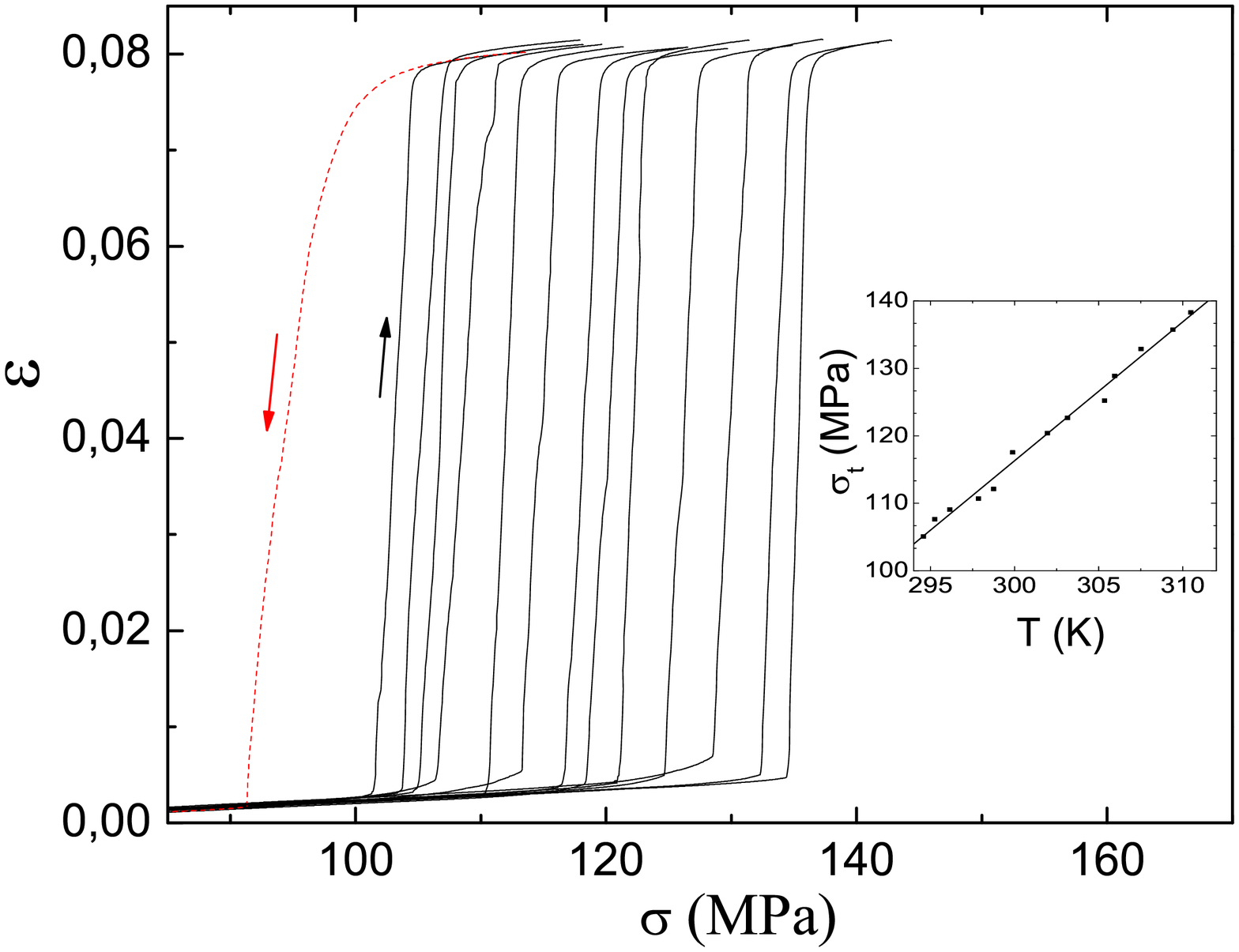}
    \caption{(Color online) Stress-strain  curves at  selected  temperatures. Continuous
    curves correspond to loading runs (from right to left) at $T =$
    310.5, 309.4, 307.5, 306.0, 305.4, 303.1, 302.0, 299.9, 298.8,
    297.9, 296.1, 295.3 and 294.6 K.  The left discontinuous curve
    corresponds to the unloading branch at $T
    =$ 294.6 K. It illustrates an example of stress-induced hysteresis
    loops.  The inset  shows the  transition stress  as a  function of
    temperature. The line is  a linear fit $\sigma  = 2.01
    (T-242)$.}  \label{Fig3}
    \end{figure}

\begin{figure}[t]
    \centering \includegraphics[width=0.75\linewidth,clip]{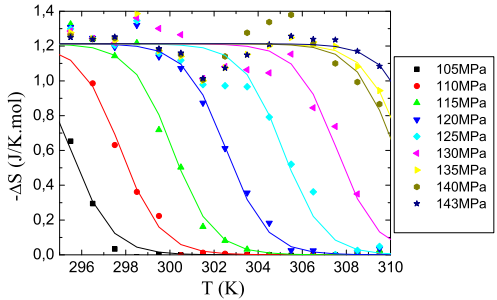}
    \caption{(Color    online)     Stress-induced    entropy    change
    (elastocaloric effect) at selected values of $\sigma$ ranging from
    105  to 143  MPa. The continuous lines are fits based on a the model
    ${\cal  F}(x)   =   \tanh^{-1}(x)$.}
    \label{Fig5}
    \end{figure}

In  Fig. \ref{Fig3} we  show  stress-strain  curves  obtained in  the stress-driven case at selected temperatures (well above the transition
temperature at  zero-stress) across the martensitic  transition in the studied Cu-Zn-Al single crystal. The shift of the transition to higher
stresses with  increasing temperature  is clearly seen.  Comparison of curves  corresponding   to  loading  and  unloading   shows  that  the
transition occurs  with weak  hysteresis of about  10 MPa.  From these curves  the elastocaloric effect  (stress-induced entropy  change) has been
obtained by  numerically  computing the  integral in  Eq. (\ref{eq1}). The stress-induced entropy change (elastocaloric effect) is shown in Fig. \ref{Fig5}.

\begin{figure}[t]
    \centering \includegraphics[width=0.75\linewidth,clip]{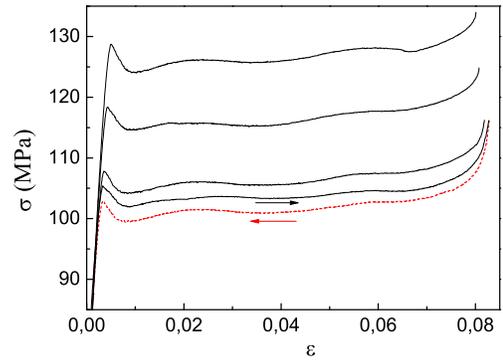}
    \caption{(Color online) Strain-stress curves obtained at selected
    temperatures. Continuous curves correspond to loading runs at
    (from top to bottom):  $T =$ 307.8, 303.1, 297.8, and 295.0 K
    The  lower discontinuous  curve corresponds to  the unloading
    branch at $T  =$ 295.0 K.  It illustrates an example of a
    strain-induced hysteresis loop.}
    \label{Fig6}
    \end{figure}

\begin{figure}[t]
    \centering \includegraphics[width=0.75\linewidth,clip]{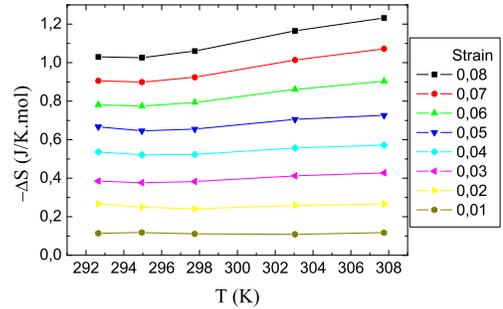}
    \caption{(Color    online)     Strain-induced    entropy    change
    (elastocaloric effect) at selected  values of $\epsilon$ from 0.01
    to  0.08.}
    \label{Fig8}
    \end{figure}

It is interesting to point out that the maximum stress-induced
entropy change  (which  corresponds  to   the  whole  entropy
change  of  the martensitic transformation) remains  almost
constant (thus independent of  temperature and  applied  stress)
over  a  very broad  temperature range \cite{comment}. This is
usually not the case for  the magnetocaloric effect in
field-induced, first-order  phase transitions for  which large
$\Delta S(T)$ is only obtained  in a  relatively narrow
temperature interval (which  depends  on  the applied  field).
Such  a difference  is  a consequence of the fact that here
tensile experiments are performed at temperatures well  above
($\geq$ 60  K) the transition  temperature at zero stress. This is
possible due  to the  strong dependence  of the transition stress
with temperature. The  upper bound is imposed by the elastic limit
of the cubic phase. By contrast, in magnetic experiments the field
is always  applied close to  the transition  temperature at zero
field. Indeed, at  higher temperatures  metamagnetic transitions
are  difficult to be  induced either  because intense  magnetic
fields that  are too  large are required or simply because  this
leads the system  above the  critical  point, where  no transition
occurs. The continuous  lines in  Fig. \ref{Fig5} are fits which
assume a shape-function ${\cal F}(x) = \tanh^{-1}(x)$. In the
strain-driven  case  we can  proceed  similarly starting  with
strain-stress curves recorded at selected temperatures. Fig.
\ref{Fig6} displays the strain-stress curves obtained at selected
temperatures. The strain-induced entropy change (elastocaloric
effect) as   a function   of   $T$    has   then been    obtained
using Eq. (\ref{eq2}). Results are shown  in Fig. \ref{Fig8}. The
variation of $\Delta  S$ with temperature reflects  a small
variation  of $\Delta \varepsilon$ with temperature.

From the  previous results the estimated \cite{comment1} entropy
change corresponding to the  whole transition  is $\Delta S_t  = -
1.20 \pm 0.15$ J/mol K consistent in both strain-  and
stress-driven cases. It is interesting to  compare this  result
with estimations  based  on the  use of the Clausius-Clapeyron
equation.    From  the  stress-driven   curves  in Fig. \ref{Fig3}
we can define a transition stress, $\sigma_t$, at each
temperature,   as  the  stress   at  the   inflection  point   of
the isotherms. The strain change,  $\Delta \varepsilon$, at the
transition can also be estimated from these curves. The transition
entropy change is  then obtained  as  $\Delta S_t  \simeq  (d
\sigma_t  / dT)  \Delta \varepsilon$.  In the  inset  of Fig.
\ref{Fig3}  we show  $\sigma_t$ vs. $T$.  We see that  $\sigma_t$
rises linearly with  increasing $T$, with a slope $d\sigma_t / dT$
= 2.01 MPa/K. Taking  an average value $\Delta \varepsilon  =
0.080 \pm 0.005$  we  obtain $\Delta S_t  = - 1.21  \pm 0.05$
J/mol  K. This value  is  compatible  with  the estimations  based
on  elastocaloric effect. It is  interesting to compare this value
with a calorimetric measurement.  To  this  end,   we  have
carried  out  calorimetric measurements  of  the   transition
entropy  change at zero stress, using a small specimen cut from
the   same original ingot. These experiments give a value $\Delta
S_t = -1.37  \pm 0.10$  J/mol K,  slightly higher (as an absolute
value) than   that  derived  from mechanical experiments. Such  a
difference should not, in principle, be attributed to
non-equilibrium effects  which have been estimated from the area
of  the hysteresis loops to be of the order  of $0.01$ J/mol K in
the stress-induced case and less than 0.001 J/mol K in
strain-induced experiments. However, it must be pointed out that
calorimetric measurements are performed in the absence of an
external field and thus a multivariant martensite is reached.
Kinematic  constraints which occur  during the transformation and
yield extra dissipative effects could explain the difference. This
is consistent with  the fact that the transition temperature in
the  multivariant  case  (234  K)  is significantly lower  than
the extrapolation to  zero stress  of the $\sigma$ vs. $T$ curve
(242 K) which corresponds to the transition to  a single  variant
martensite (see inset of Fig. \ref{Fig3}).

The  elastocaloric  effect  has  been  reported  in  the  Fe-Rh
alloy \cite{Nikitin1992}. In this case  the entropy change has a
magnetic origin associated with  a re-orientation of spins taking
place at the first-order   antiferromagnetic  to   ferromagnetic
magnetostructural transition which is induced by the application
of a tensile stress due to  strong  magnetoelastic   coupling.
Similar  effects  (termed  the barocaloric effect) have been
reported  for a  number  of rare-earth compounds subjected  to
uniaxial pressure  \cite{Strassle2003}. In the present study,
however, the elastocaloric effect is associated with a purely
structural transition.  In this  case, the  entropy  change is
predominantly vibrational and originates  from the  very low
energy TA$_2$ transverse  ([110] propagation and  [1 $\bar{1}$0]
polarization) phonon branch of  the cubic phase \cite{Planes2001}.
It  is also worth noticing that in Fe-Rh the  entropy change is
positive when the stress is  isothermally applied, and thus  the
sample cools  down when  the stress  is  adiabatically  applied.
This corresponds  to  an  inverse elastocaloric effect which is
the analog of the inverse magnetocaloric effect      reported
in      Heusler      martensitic      alloys \cite{Krenke2005}.
Notice that inverse effects are only possible when there is
strong coupling between  magnetic and structural  degrees of
freedom.

The  adiabatic temperature  change associated  with  the elastocaloric effect  can  be  estimated as  $\Delta  T  \simeq  - \frac{T}{C}  \Delta  S
\label{DeltaT}$, where $C$  is the specific heat which  is assumed to be stress  independent. For  Cu-Zn-Al, the  value for  the  specific heat
close to  room temperature  is  approximately  25  J/K mol  in  both martensitic  and  cubic phases  \cite{Lashley2007}.  For an  adiabatic stress
drop of about  10 MPa (for instance from 110 MPa  to 100 MPa at 300K, see Fig. \ref{Fig3})  the maximum expected temperature change is 15K. Notice
that this  value is orders  of magnitude larger  than the typical values  in elastic solids  far from any phase  transition. The effect  of  strain
rate on  the martensitic  transition  of  several Cu-based  shape-memory   alloys  was   studied  by  means   of  direct measurements   of  the
temperature  changes   associated   with  the transition.    For   Cu-Al-Ni   \cite{Rodriguez1980}    and   Cu-Zn-Sn \cite{Brown1981} crystals (with
the  same martensitic structure as our Cu-Zn-Al  crystal), the  measured temperature  changes at  high strain rates (close to the adiabatic limit)
are 14 K and 12 K, respectively, for a strain  rate of 25 min$^{-1}$. These  temperature values compare well to the value indirectly  computed here
for Cu-Zn-Al. On the other hand, present values for Cu-Zn-Al are comparable to those reported for other elastocaloric materials undergoing
first-order magnetostructural phase transitions: maximum changes of 8.7  K and 14 K are computed for Fe-Rh   \cite{Nikitin1992}  and Eu-Ni-Si-Ge
\cite{Strassle2003}, respectively.

The elastocaloric effect associated with the martensitic transition in a  Cu-Zn-Al single  crystal has  been studied. It is formally equivalent  to
the  magnetocaloric  effect in  a magnetic  system and in  our  case it  should  be  comparable to the magnetocaloric effect  in the  vicinity of a
first-order metamagnetic transition.  However,  while   the  magnetocaloric effect  is  always determined by the field  inducing the metamagnetic
transition, for the mechanical case it has been possible to obtain the isothermal entropy change  by inducing the structural transition  using both
strain and stress.  While the transition  path is  essentially different  in both cases, the corresponding isothermal entropy  changes are the  same
to within errors. This  is  due to  the  fact that  hysteresis is  small independent of  the driving  mechanism. For practical applications of
caloric effects, the refrigerant capacity is a central parameter to be considered  \cite{Gschneidner2005}.  It is defined as  ${\cal  R}  =
\int_{\Delta T}  \Delta S(T) dT \simeq \Delta S \Delta T =  - \Delta \varepsilon  \Delta \sigma$, where $\Delta \sigma$  is the  change of $\sigma$
necessary to change the transition temperature by $\Delta T$ ($\Delta  S$  and $\Delta \varepsilon$  are  assumed constants).  The interest in our
case is that  $\Delta \sigma$ can be chosen in a broad range of values which   opens  up  interesting   opportunities  in refrigeration applications
based on the elastocaloric effect.

This work  received financial support from CICyT  (Spain), Project No. MAT2007-61200,   Marie-Curie   RTN   MULTIMAT   (EU), Contract No.
MRTN-CT-2004-505226,      and      DURSI (Catalonia), Project No. 2005SGR00969. The authors acknowledge O. Toscano for experimental assistance.


\begin{thebibliography}{99}
\bibitem{comment} The isothermal entropy change will significantly decrease below the martensitic transition temperature at
zero stress.
\bibitem{comment1} Error limits are based on reproducibility of
experiments.
\end{thebibliography}
\end{document}